# Estimating Box-Cox power transformation parameter via goodness of fit tests

Özgür Asar[a,*], Özlem İlk[b], Osman Dağ[b]

[a] CHICAS, Lancaster Medical School, Lancaster University, Lancaster, United Kingdom
[b] Department of Statistics, Middle East Technical University, Ankara, Turkey

Box-Cox power transformation is a commonly used methodology to transform the distribution of a non-normal data into a normal one. Estimation of the transformation parameter is crucial in this methodology. In this study, the estimation process is hold via a searching algorithm and is integrated into well-known seven goodness of fit tests for normal distribution. An artificial covariate method is also included for comparative purposes. Simulation studies are implemented to compare the effectiveness of the proposed methods. The methods are also illustrated on two different real life data applications. Moreover, an R package **AID** is proposed for implementation.

## 1. Introduction

Normal distribution has a fundamental role in statistical literature, since it forms the basis of most of the statistical methods such as regression analysis, analysis of variance and t-test. Therefore, the validity of the related results necessitates the agreement between the distribution of the observed data and this theoretical distribution. In cases where this agreement is deteriorated, which is common in real life datasets, transformation methods might be a practical remedy to secure it. The most popular and commonly used method is the Box-Cox power transformation (Box and Cox, 1964). Since its proposition, it has been applied in various fields. Some of the recent works include Lee et al. (2013), Gillard (2012) and Sun et al. (2011). Box-Cox transformation mainly applies a deterministic power function to the raw data by using the estimate of the power transformation parameter, $\lambda$. Therefore, the estimation of $\lambda$ is crucial. The original proposal of the methodology (Box and Cox, 1964) involved the maximum likelihood estimation (MLE). Alternative methodologies included the works of Rahman (1999), Rahman and Pearson (2013), and Dag et al. (2013). Whereas the first two studies proposed the estimation of $\lambda$ via two normality tests, specifically Shapiro-Wilk and Anderson-Darling tests, respectively, the third one proposed simulating a single artificial and non-informative covariate and finding $\hat{\lambda}$ which minimizes the sum of squared error among several simple linear regression models. These studies showed that the MLE of $\lambda$ might be biased and inefficient.

    Instead of using the MLE, the works of Rahman (1999) and Rahman and Pearson (2008) used Newton-Raphson (N-R) algorithm to obtain $\hat{\lambda}$ within the aforementioned normality tests. However, it is well known that this procedure has some disadvantages. For instance, the method requires an initial point selection and it might capture a nearest root, i.e., a local root, instead of the global one. Moreover, its application to other goodness of fit tests, such as Lilliefors test, is challenging. Besides, there is no user-friendly software to estimate $\lambda$ with the

---

aforementioned goodness of fit tests by using the N-R procedure. The MATLAB (2010) codes of Rahman (1999) and Rahman and Pearson (2008) are available from the original authors for Shapiro-Wilk and Anderson-Darling tests, but they are not in the form of public use. Moreover, these codes are restricted only to specific sample sizes: 20, 40 and 100.

In this study, we extended their work in four perspectives: 1) We considered the estimation of λ with a different methodology, specifically a searching algorithm which finds the argument of maximum (arg max) or minimum (arg min) over a pre-specified interval in which candidate λ values lay; 2) We utilized the estimation via seven well-known goodness of fit tests and via a new method by Dag et al. (2013); 3) We implemented a publicly available R [15] package **AID**; 4) This package was not restricted to any sample size choices.

We conducted two simulation studies. The first one questioned whether our searching method was at least as good as the numerical root finding methods such as N-R algorithm to estimate λ. The second simulation study was conducted to evaluate the performances of all the methods including the artificial covariate method of Dag et al. (2013) under different conditions. Furthermore, methods were illustrated on two different real life datasets with different characteristics; one of them was right skewed and the other was left skewed.

The organization of the paper is as follows. In Section 2, we provide brief details of Box-Cox power transformation and discuss the parameter estimation methodologies. The results of the simulation studies are provided in Section 3. Real life applications and related results are introduced in Section 4. We discuss the implementation of the methods via the package **AID** and some computational aspects in this Section as well, and close the article by conclusions in Section 5.

## 2. Methods

The Box-Cox power transformation [5] on observations $y_i$ (i=1,2,…,n) is given by

$$y_i^{(\lambda)} = \begin{cases} \frac{y_i^\lambda - 1}{\lambda}, & \text{if } \lambda \neq 0 \\ \log y_i, & \text{if } \lambda = 0 \end{cases}, \quad (1)$$

where λ is the unknown power transformation parameter and n is the sample size. Generally, the transformation given in Equation 1 is known as the conventional Box-Cox transformation. Nonetheless, due to the fact that the analysis of variance does not change by linear transformation, an alternative version is usually considered:

$$y_i^{(\lambda)} = \begin{cases} y_i^\lambda, & \text{if } \lambda \neq 0 \\ \log y_i, & \text{if } \lambda = 0 \end{cases}, \quad (2)$$

In this study, we achieved the estimation of λ via well-known seven goodness of fit tests for normality; namely, Shapiro-Wilk, Anderson-Darling, Cramer-von Mises, Pearson Chi-square, Shapiro-Francia, Lilliefors (Kolmogorov-Smirnov) and Jarque-Bera tests and via a new artificial covariate method by Dag et al. (2013). Brief information on these tests and artificial covariate method can be found in the Appendix. Interested readers are referred to Noughabi and Arghami (2011) and Yap and Sim (2011) for power comparisons of some of these tests through Monte Carlo simulation. We preferred using searching algorithms to obtain the optimum value, i.e., arg max for Shapiro-Wilk and Shapiro-Francia tests and arg min for the rest of the normality tests, to find the estimate of λ. The algorithm is provided in the following steps:

Our proposed algorithm for the estimation of λ

i) Select a sequence of candidate λ values by a fairly precise increment such as 1/10, 1/50 and so on, in a specified interval.
ii) Be sure that each element of the dataset is positive. Otherwise, add a small constant to all observations to shift the location to positive values (as originally proposed by Box and Cox [5]).
iii) Apply Box-Cox power transformation given in Equation 1 by using all of the candidate λ values and obtain several transformed samples, as many as the λ values.
iv) Check the normality of each of these transformed samples by the specified goodness of fit tests.
v) Select the λ value for which the maximum or minimum test statistic, depending on the choice of the test, is obtained.
vi) Control whether this λ value is able to satisfy the normality of the transformed data by three of the aforementioned seven normality tests. Furthermore, check whether it yields global maximum or minimum by graphical analysis.
vii) If these checks do not support the selection of the candidate λ value, then enlarge the range of the lambda sequence and repeat steps ii) to vi) again.

We used three goodness of fit tests in order to check the normality in the validation step after the transformation; namely, Shapiro-Wilk test, Shapiro-Francia test and Jarque-Bera test. The question of why we used these three goodness of fit tests in validation stage of normality might arise. Within a group that is composed of Shapiro-Wilk test, Anderson-Darling test, Lilliefors test, Cramer-von Mises test, Jarque-Bera test and Pearson Chi-square test; Shapiro-Wilk test is the best one for the asymmetric distributions and is powerful for symmetric short tailed distributions (Yap and Sim, 2011). Shapiro-Francia is the modification of Shapiro-Wilk test for large sample sizes (See Appendix for the details). We also added Jarque-Bera test in the validation part since it is superior to Shapiro-Wilk test when data have symmetric distribution with medium or long tails or slightly skewed distribution with long tails (Thadewald and Buning, 2007). Hence, we considered the validation step for different possible types of data.

For the validation of normality after transformation, it is better to adjust the p-values for multiple comparisons since we have used more than one normality test. Benjamini and Hochberg (1995) suggested a simple procedure to control the false discovery rate. It is more powerful than the procedures controlling the traditional familywise error rate for independent test statistics. The same procedure was verified in Benjamini and Yekutieli (2001) to control the false discovery rate when positive dependency of test statistics existed. Herewith, it was set as default in the function boxcoxnc proposed under the R package **AID**.

The package **AID** to implement our methodologies is available from Comprehensive R Archive Network (CRAN) at http://CRAN.R-project.org/package=AID.

## 3. Results on simulation studies

Two simulation studies are carried out to answer different questions. In the next two subsections, these questions, algorithm and results of the simulation studies are provided.

### 3.1. Simulation study I

In this subsection, we conducted a simulation study to answer the following question: "Do our proposed estimation methodology, i.e., searching algorithm, work at least as good as the ones of Rahman (1999) and Rahman and Pearson (2008)?". Our simulation study included different scenarios considered in these two works, specifically, different sample sizes, and different combinations of mean, standard deviation and true $\lambda$ values. The related methodology could be depicted in the steps below.

Algorithm for simulation study I

i) Generate a random sample from normal distribution with mean μ (μ = $-15, -10, -5, 5, 10, 15$) and standard deviation σ (σ = 1, 2) with a sample size n (n=20, 100) by using the MATLAB code of Rahman and Pearson (2008).
ii) Apply inverse Box-Cox transformation defined by $(y\lambda + 1)^{1/\lambda}$ for $\lambda = -2, -1, -0.5, 0.5, 1, 2$) to create non-normal samples.
iii) Estimate λ by the methods of Rahman (1999) and Rahman and Pearson (2008) by utilizing the MATLAB code.
iv) Extract the datasets generated in MATLAB to a file and read them into R.
v) Estimate λ by our proposed methodology.

Following Rahman (1999) and Rahman and Pearson (2008), the replication number is selected as 1,000. Standard accuracy measures such as mean, bias, percentage bias, standard error (SE) and mean squared error (MSE) were calculated to assess the comparison. However, only the bias, SE and MSE are reported here due to page limitations.

The N-R algorithm reached convergence within small number of iterations. For instance, the algorithm, that used the N-R, converged in at most 4 steps, even for small sample sizes such as 20, with a convergence threshold of $10^{-5}$. The initial value of the transformation parameter in all Newton processes was chosen as 1.

Results of our method are presented, under the heading of O, together with the results of Newton-Raphson method, under the heading of N, in Table 1. Our method yielded smaller or similar bias and MSEs compared to the N-R procedure for both Shapiro-Wilk and Anderson-Darling tests under most sample sizes and mean, standard deviation combinations. The main differences between the two methods were observed when the absolute value of λ gets larger for small sample size. For instance, the MSE for our method was observed to be 5.877 [6.454] for Shaprio-Wilk [Anderson-Darling] test, whereas it was found to be 7.163 [8.176] with their procedure when n=20 under μ= -5, σ=1 and $\lambda = -2$. For smaller values of λ these two methods seemed to be working similarly. For n=100, the results for these two methods were similar even for $\lambda = -2$ and 2. In the light of these results, we can conclude that our searching approach performs at least as good as the available procedure, i.e., N-R method, for Shapiro-Wilk and Anderson-Darling tests in terms of estimating the Box-Cox power transformation parameter, λ, regarding both consistency and efficiency.

[Insert Table 1 here]

### 3.2. Simulation study II

Next, we conducted a simulation study to answer the following questions: "Which one of the eight methods does yield better results?" and "Are there any differences in the results based on the sample size?" The methodology followed in this simulation study is provided in the following steps.

Algorithm for simulation study II

i) Generate a random sample from normal distribution with mean μ (μ = 0) and standard deviation σ (σ = 1, 5) for sample size n (n=20, 30, 50, 100, 500).
ii) Apply inverse transformation $(y)^{1/\lambda}$ by considering the Box-Cox transformation defined in Equation 2 with power transformation parameter, λ (λ= -5, -2, -1, 0, 2, 5).
iii) Estimate λ by using our proposed approaches.

The simulation was repeated for 10,000 runs. Results are presented in Table 2. For this simulation study, we only reported the bias and MSE, despite the fact that other statistics such as mean, percentage bias and standard error were also calculated.

As expected, biases and MSEs became smaller as the sample size increased and/or the magnitude of λ decreased. Almost all of the estimation approaches performed similar to each other, especially when the sample size increased. However, Pearson Chi-square test showed less consistent and efficient performance at small sample sizes. For instance, under n=20 and λ= -5, the bias and MSE were found to be -0.756 and 6.241, respectively, whereas the related values for other methods such as Shapiro-Wilk were found to be 0.039 and 3.323. As the magnitude of λ decreased and the sample size increased, its inefficient performance disappeared. In fact, all of the methods, including Chi-square method, performed well at λ=0 regardless of the parameter combinations (σ, n) in data generation. To illustrate, the biases and MSEs of all the tests were practically 0 for n=500 and λ=0. The artificial covariate method performed as the most consistent method especially for small sample sizes and large λ values in the magnitude, although sizable biases were observed for it. In fact, it yielded the smallest MSEs under all conditions. Moreover, as the sample size increased, its biasness tended to disappear.

When the standard deviation of the data generation distribution was decreased, all of the methods showed higher inconsistencies through higher MSEs; though estimation with the artificial covariate method was less biased. For instance, under N(0, 25) with n=20 and λ=-5, the MSE was found to be 4.639 for the Lilliefors test; it increased to 7.472 under N(0, 1) when the other parameters were kept same. Nevertheless, since the comparison of methods remained same regardless of σ, only the results for N(0,25) were included here due to space limitations. Results for N(0,1) are available as Supplementary Data.

Within the techniques discussed in this study, artificial covariate approach was the best one with respect to the MSE criteria. Moreover, this approach was found to be superior compared to MLE for estimating λ in Dag et al. (2013). The biases obtained with artificial covariate method seemed to be slightly higher than the other biases. However, the bias of an estimate was stated to be within the acceptable range if it was less than its SE/2; some references even stated this value as 2SE (Sinharay et al., 2001, Burton et al., 2006). All methods discussed in this study, including artificial covariate method, had small biases, even smaller than SE/2. The method utilizing Shapiro-Wilk test also provided better results compared to the methods using other normality tests with respect to bias and MSE in most of the cases. To illustrate, it was noted that the bias and MSE values were the smallest for the sample size of 50 when Shapiro-Wilk test was utilized to estimate λ for all true λ combinations when data were generated from N(0, 25).

In brief, Pearson Chi-square test might not be preferred to estimate λ. Other methods could be studied together to obtain a consensus perspective. However, artificial covariate method and Shapiro-Wilk test seem to be more effective than most of the tests.

[Insert table 2 here]

## 4. Results on real data

Our methods are illustrated on two real life data applications in this section. Brief information on datasets is provided together with results. Some computational details are discussed in the last subsection.

### 4.1. Data on textile

Data exerted in this application were collected by International Wool Textile Organization to detect the impact of some factors on the number of cycles to failure of worsted yarn [5]. The factors described in that experiment were the length of test specimen (250, 300, 350 mm.), amplitude of loading cycle (8, 9, 10 mm.), and load (40, 45, 50 gm.). Data involved only 27 observations. Data are available in the R package **BHH2** (Barrios, 2012) under the name of ***woolen.data***.

Density plot of cycles to failure indicated right-skewness of the data with possibly a mixture of two densities (Figure 1, left panel). Moreover, all of the normality tests pointed out that there was enough evidence to conclude that cycles to failure followed a non-normal distribution (e.g., p-value for Shapiro-Wilk $=3.031\times 10^{-5}$).

The estimates of transformation parameter under all eight methods are presented in Table 3. Moreover, the p-values of three normality tests on transformed datasets are provided in this table. For instance, Shapiro-Wilk test yielded a p-value of 1.00 when applied to the transformed data with the $\hat{\lambda}$ obtained by Anderson-Darling test. All of these p-values indicated that there was not enough evidence towards non-normality of these transformed samples. Moreover, density plots of all of the transformed cycles to failure demonstrated symmetric distributions around their means (Figure 1, right panel). Note that Box-Cox (Box and Cox, 1964) suggested using $\hat{\lambda} = -0.06$ which was supported exactly by four of our methods; namely, Shapiro-Wilk, Shapiro-Francia, Lilliefors and Jarque-Bera tests. The estimates obtained by the other methods were approximately equal to this original estimate; these estimates were also successful in terms of converting the distribution to a normal one. Note that the Pearson chi-square result seemed to be the most adverse one.

[Insert table 3 here]

[Insert figure 1 here]

### 4.2. Data on students' grades

Data used in this application included grades of 42 college students collected at Middle East Technical University of Turkey. The data were exerted to illustrate the behavior of the approaches proposed in this paper on negatively skewed data (figure is not shown here).

All of the tests suggested non-normality of the raw data (e.g., Shapiro-Wilk test p-value $=3.237\times 10^{-3}$). The proposed methods were implemented and the estimates of power transformation parameters and the results of normality tests on the transformed datasets are displayed in Table 4. The p-values suggested that the transformations to normality were successful.

[Insert table 4 here]

### 4.3. Implementation

The function boxcoxnc under the R package **AID** was proposed to implement all of the aforementioned methods. The estimates of λ's for the textile and students' grades datasets can be obtained by the following R codes:

```
R> install.packages("AID")
R> library(AID)
R> data(textile)
R> boxcoxnc(textile$textile )
R> data(grades)
R> boxcoxnc(grades$grades)
```

The computational times are provided in Table 5 for both real life datasets. Artificial covariate method demands slightly longer times since its reliable implementation requires repetitions. Nevertheless, the methods took less than 3 seconds even when all of the methods were run together. The analyses were done on a PC with 6.00 GB RAM and 2.50 GHz processor.

[Insert table 5 here]

## 5. Conclusion

In this study, different approaches were proposed to estimate Box-Cox power transformation parameter. Our proposed methods were based on different normality tests and we used searching algorithms to find maximum or minimum instead of numerical methods such as Newton-Raphson.

We conducted two simulation studies. First one was based on the comparison of searching algorithms and numerical methods, and the related results indicated that the former one was performing at least as good as the N-R root finding procedure. The second study was based on evaluating the features of the proposed methods under different scenarios. Results showed that artificial covariate method and Shapiro-Wilk test seem to be more effective than most of the tests in attaining the true transformation parameter. It was observed that Pearson Chi-square test was able to compete with the other methods only for large sample sizes. The methods were applied on two real life datasets. We proposed an R package, **AID,** for implementation.

**Acknowledgement**

This work received best paper award among papers presented in "y-BIS 2013: Joint Meeting of Young Business and Industrial Statisticians" held on 19-21 September, 2013.

**Appendix**

*Shapiro-Wilk & Shapiro-Francia tests*

Shapiro and Wilk (1965) proposed a goodness of fit test, which was named after its founders, and was specifically designed to test the normality of a dataset. The calculation of the Shapiro-Wilk test statistic is defined by

$$W = \frac{b^2}{((n-1)s^2)} \quad , \tag{A1}$$

$$b = \sum_{i=1}^{n/2} a_{n-i+1}(x_{(n-i+1)} - x_{(i)}),$$

where $a = (w'V^{-1})/\sqrt{(w'V^{-1}V^{-1}w)}$, V is the covariance matrix of order statistics, w is the vector of expected value of order statistics, and $x_{(i)}$ are sample order statistics [24].

Alternatively, Shapiro and Francia (1972) pointed out that order statistics for large samples behaved independently and they proposed a modification of the Shapiro-Wilk test statistic for large sample sizes. The Shapiro-Francia test statistic is defined by

$$W' = \frac{(a^*x)^2}{((n-1)s^2)}, \tag{A2}$$

where $a^* = (w')/\sqrt{(w'w)}$. Weisberg and Bingham (1975) suggested an approximation to $w_i$ defined by $w_i = \Phi^{-1}(\frac{i-0.375}{n+0.25})$, $\Phi^{-1}$ is the inverse of cumulative distribution function of standard normal distribution.

Both Shapiro-Wilk and Shapiro-Francia test statistics take the maximum value of 1, which indicates less evidence towards non-normality.

*Anderson-Darling test*

Anderson and Darling (1954) presented an empirical distribution function (EDF) test for normality. The related test statistic is given by

$$A^2 = -n - n^{-1}\sum_{i=1}^{n}[2i-1][\ln(p_{(i)}) + \ln(1-p_{(n-i+1)})], \tag{A3}$$

where $p_{(i)} = \Phi([x_{(i)} - \bar{x}]/s)$. The values of the Anderson-Darling test statistic which are close to 0 indicate less evidence towards non-normality.

*Cramer-von Mises test*

The Cramer-von Mises test statistics (1928) is given by

$$W^2 = \frac{1}{12n} + \sum(p_{(i)} - \frac{2i-1}{2n})^2. \tag{A4}$$

As Cramer-von Mises test statistic is adjacent to the value of 0, which is the minimum value of the test statistic, it indicates less evidence towards non-normality.

*Pearson Chi-square test*

The Pearson Chi-Square test (Pearson, 1900) statistic is described by

$$\chi^2 = \sum_{i=1}^{k} \frac{(n_i - np_i)^2}{np_i}, \tag{A5}$$

where n observations are separated into k exclusive classes, $p_i$ is the probability of an observation to appear in class i under null hypothesis, $n_i$ is the number of observations in the $i^{th}$ class (Thode, 2002). Larger values of the test statistic indicate departures from normality.

*Lilliefors (Kolmogorov-Smirnov) test*

Lilliefors (1967) is an EDF based goodness of fit test for normality and the test statistic depends on maximum difference between the empirical and hypothetical cumulative distribution functions. The test statistic is given by

$$D^+ = \max_{i=1,\dots,n}\left[\frac{i}{n} - p_{(i)}\right], D^- = \max_{i=1,\dots,n}\left[p_{(i)} - \frac{i-1}{n}\right]$$

$$D = \max[D^+, D^-] \tag{A6}$$

where $p_{(i)} = \Phi([x_{(i)} - \bar{x}]/s)$. The values of the Lilliefors test statistic closer to 0 indicate less evidence towards non-normality.

*Jarque-Bera test*

Jarque and Bera (1987) proposed a skewness and kurtosis based goodness of fit test for the normality and the related test statistic is

$$LM = \frac{n}{6}\left[S^2 + \frac{(K-3)^2}{4}\right], \tag{A7}$$

where S and K are skewness and kurtosis, respectively. The values of the test statistic close to 0 indicate less evidence towards non-normality.

*Method of artificial covariate*

Dag et al. (2013) proposed simulating a single non-informative covariate from normal distribution with mean 0 and standard deviation 100 when no covariate was available. The usual Box-Cox power transformation parameter was applied by using this artificial covariate. This methodology was included in this study to compare its performance in terms of estimating λ with the proposed goodness of fit test methodologies.

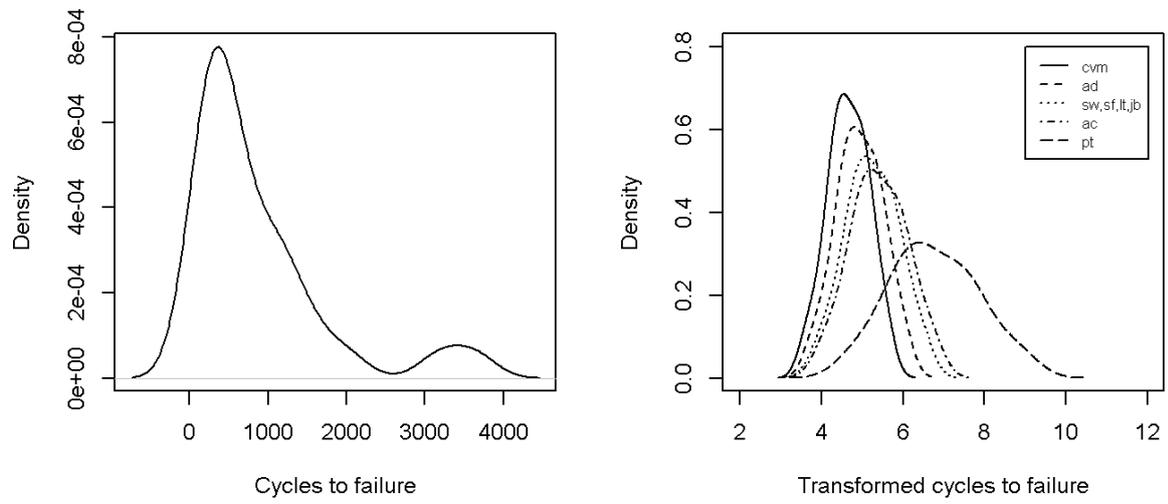

**Figure 1.** Density plot of cycles to failure before (left) and after transformations (right).

**Table 1. Bias, SE and MSE estimates of λ**

| n | μ | σ | λ | | O-SW | N-SW | O-AD | N-AD |
|---|---|---|---|---|---|---|---|---|
| 20 | -5 | 1 | -2 | Bias | 0.111 | 0.128 | 0.061 | 0.071 |
| | | | | SE | 2.422 | 2.673 | 2.540 | 2.858 |
| | | | | MSE | 5.877 | 7.163 | 6.454 | 8.176 |
| | -10 | 2 | -2 | Bias | 0.114 | 0.130 | 0.065 | 0.072 |
| | | | | SE | 2.338 | 2.546 | 2.458 | 2.723 |
| | | | | MSE | 5.478 | 6.498 | 6.044 | 7.419 |
| | -5 | 1 | -1 | Bias | 0.063 | 0.064 | 0.036 | 0.035 |
| | | | | SE | 1.449 | 1.460 | 1.548 | 1.563 |
| | | | | MSE | 2.105 | 2.137 | 2.399 | 2.443 |
| | -10 | 2 | -1 | Bias | 0.063 | 0.064 | 0.036 | 0.036 |
| | | | | SE | 1.330 | 1.337 | 1.422 | 1.429 |
| | | | | MSE | 1.772 | 1.791 | 2.023 | 2.044 |
| | -10 | 1 | -0.5 | Bias | 0.043 | 0.035 | 0.019 | 0.017 |
| | | | | SE | 1.492 | 1.495 | 1.561 | 1.578 |
| | | | | MSE | 2.228 | 2.235 | 2.438 | 2.491 |
| | -15 | 2 | -0.5 | Bias | 0.037 | 0.031 | 0.018 | 0.017 |
| | | | | SE | 1.055 | 1.041 | 1.109 | 1.115 |
| | | | | MSE | 1.115 | 1.085 | 1.231 | 1.243 |
| | 10 | 1 | 0.5 | Bias | -0.037 | -0.032 | -0.016 | -0.018 |
| | | | | SE | 1.479 | 1.475 | 1.560 | 1.577 |
| | | | | MSE | 2.188 | 2.178 | 2.434 | 2.488 |
| | 15 | 2 | 0.5 | Bias | -0.038 | -0.031 | -0.017 | -0.018 |
| | | | | SE | 1.055 | 1.043 | 1.108 | 1.114 |
| | | | | MSE | 1.115 | 1.088 | 1.229 | 1.242 |
| | 5 | 1 | 1 | Bias | -0.064 | -0.065 | -0.034 | -0.036 |
| | | | | SE | 1.453 | 1.461 | 1.546 | 1.559 |
| | | | | MSE | 2.114 | 2.140 | 2.390 | 2.433 |
| | 10 | 2 | 1 | Bias | -0.065 | -0.066 | -0.035 | -0.036 |
| | | | | SE | 1.333 | 1.337 | 1.421 | 1.429 |
| | | | | MSE | 1.782 | 1.792 | 2.021 | 2.044 |
| | 5 | 1 | 2 | Bias | -0.104 | -0.132 | -0.051 | -0.072 |
| | | | | SE | 2.430 | 2.674 | 2.535 | 2.859 |
| | | | | MSE | 5.916 | 7.166 | 6.430 | 8.177 |
| | 10 | 2 | 2 | Bias | -0.105 | -0.130 | -0.052 | -0.073 |
| | | | | SE | 2.347 | 2.548 | 2.451 | 2.726 |
| | | | | MSE | 5.518 | 6.509 | 6.008 | 7.435 |

Note: O-SW and O-AD were obtained by our proposed approach, while N-SW and N-AD were obtained by Newton-Raphson root finding procedure. Shapiro-Wilk and Anderson-Darling test statistics were used in Box-Cox transformation for both methods.

**Table 1. (Continuation)**

| n | μ | σ | λ | | O-SW | N-SW | O-AD | N-AD |
|---|---|---|---|---|---|---|---|---|
| | -5 | 1 | -2 | Bias | 0.029 | -0.151 | -0.013 | -0.005 |
| | | | | SE | 0.885 | 1.042 | 0.995 | 1.060 |
| | | | | MSE | 0.784 | 1.108 | 0.990 | 1.124 |
| | -10 | 2 | -2 | Bias | 0.031 | -0.151 | -0.011 | -0.004 |
| | | | | SE | 0.847 | 0.990 | 0.959 | 1.009 |
| | | | | MSE | 0.719 | 1.003 | 0.919 | 1.018 |
| | -5 | 1 | -1 | Bias | 0.018 | -0.075 | -0.003 | -0.003 |
| | | | | SE | 0.504 | 0.571 | 0.581 | 0.581 |
| | | | | MSE | 0.254 | 0.332 | 0.337 | 0.338 |
| | -10 | 2 | -1 | Bias | 0.018 | -0.075 | -0.003 | -0.002 |
| | | | | SE | 0.459 | 0.521 | 0.530 | 0.530 |
| | | | | MSE | 0.211 | 0.277 | 0.281 | 0.281 |
| | -10 | 1 | -0.5 | Bias | 0.010 | -0.035 | -0.004 | -0.004 |
| | | | | SE | 0.515 | 0.584 | 0.590 | 0.591 |
| | | | | MSE | 0.265 | 0.342 | 0.348 | 0.349 |
| 100 | -15 | 2 | -0.5 | Bias | 0.010 | -0.037 | -0.003 | -0.003 |
| | | | | SE | 0.362 | 0.411 | 0.416 | 0.416 |
| | | | | MSE | 0.131 | 0.170 | 0.173 | 0.173 |
| | 10 | 1 | 0.5 | Bias | -0.005 | 0.045 | -0.005 | -0.005 |
| | | | | SE | 0.515 | 0.584 | 0.589 | 0.590 |
| | | | | MSE | 0.265 | 0.343 | 0.347 | 0.348 |
| | 15 | 2 | 0.5 | Bias | -0.006 | 0.043 | -0.004 | -0.004 |
| | | | | SE | 0.362 | 0.411 | 0.415 | 0.415 |
| | | | | MSE | 0.131 | 0.171 | 0.172 | 0.172 |
| | 5 | 1 | 1 | Bias | -0.014 | 0.085 | -0.005 | -0.005 |
| | | | | SE | 0.504 | 0.573 | 0.579 | 0.579 |
| | | | | MSE | 0.254 | 0.336 | 0.335 | 0.335 |
| | 10 | 2 | 1 | Bias | -0.014 | 0.084 | -0.005 | -0.005 |
| | | | | SE | 0.459 | 0.522 | 0.527 | 0.528 |
| | | | | MSE | 0.211 | 0.280 | 0.278 | 0.279 |
| | 5 | 1 | 2 | Bias | -0.029 | 0.168 | -0.012 | -0.010 |
| | | | | SE | 0.892 | 1.045 | 0.997 | 1.056 |
| | | | | MSE | 0.797 | 1.121 | 0.994 | 1.115 |
| | 10 | 2 | 2 | Bias | -0.029 | 0.168 | -0.012 | -0.010 |
| | | | | SE | 0.854 | 0.994 | 0.960 | 1.004 |
| | | | | MSE | 0.731 | 1.017 | 0.921 | 1.009 |

Note: O-SW and O-AD were obtained by our proposed approach, while N-SW and N-AD were obtained by Newton-Raphson root finding procedure. Shapiro-Wilk and Anderson-Darling test statistics were used in Box-Cox transformation for both methods.

**Table 2. Comparison of bias, SE and MSE estimates of λ with 8 different methods when data is generated from N(0, 25)**

| N | true λ | | SW | AD | CVM | PT | SF | LT | JB | AC |
|---|---|---|---|---|---|---|---|---|---|---|
| 20 | -5 | Bias | 0.039 | -0.079 | -0.101 | *-0.756* | -0.112 | **-0.015** | 0.087 | 0.624 |
| | | SE | 1.822 | 1.933 | 2.070 | *2.381* | 1.845 | 2.154 | 1.909 | **1.485** |
| | | MSE | 3.323 | 3.744 | 4.297 | *6.241* | 3.415 | 4.639 | 3.653 | **2.594** |
| | -2 | Bias | -0.008 | -0.065 | -0.086 | *-0.554* | -0.073 | -0.068 | **0.000** | 0.250 |
| | | SE | 0.802 | 0.873 | 0.960 | *1.423* | 0.820 | 1.037 | 0.876 | **0.600** |
| | | MSE | 0.643 | 0.767 | 0.929 | *2.331* | 0.677 | 1.079 | 0.768 | **0.423** |
| | -1 | Bias | -0.004 | -0.033 | -0.044 | *-0.295* | -0.037 | -0.037 | **-0.003** | 0.126 |
| | | SE | 0.407 | 0.449 | 0.499 | *0.772* | 0.417 | 0.535 | 0.463 | **0.302** |
| | | MSE | 0.166 | 0.203 | 0.251 | *0.683* | 0.175 | 0.288 | 0.214 | **0.107** |
| | 0 | Bias | -0.001 | -0.001 | -0.001 | *-0.017* | -0.001 | -0.001 | -0.001 | **0.000** |
| | | SE | **0.045** | 0.055 | 0.055 | *0.082* | 0.055 | 0.063 | 0.055 | **0.045** |
| | | MSE | **0.002** | 0.003 | 0.003 | *0.007* | 0.003 | 0.004 | 0.003 | **0.002** |
| | 2 | Bias | **0.001** | 0.058 | 0.080 | -0.166 | 0.067 | 0.064 | -0.005 | *-0.254* |
| | | SE | 0.802 | 0.875 | 0.967 | *1.273* | 0.820 | 1.048 | 0.879 | **0.598** |
| | | MSE | 0.643 | 0.769 | 0.941 | *1.649* | 0.677 | 1.103 | 0.773 | **0.422** |
| | 5 | Bias | -0.045 | 0.072 | 0.096 | *-0.750* | 0.106 | **0.014** | -0.101 | -0.634 |
| | | SE | 1.843 | 1.956 | 2.086 | *2.248* | 1.864 | 2.163 | 1.918 | **1.491** |
| | | MSE | 3.398 | 3.832 | 4.362 | *5.617* | 3.486 | 4.678 | 3.689 | **2.624** |
| 30 | -5 | Bias | **-0.008** | -0.118 | -0.131 | *-0.554* | -0.149 | -0.064 | 0.053 | 0.445 |
| | | SE | 1.560 | 1.692 | 1.838 | *2.249* | 1.584 | 1.940 | 1.622 | **1.297** |
| | | MSE | 2.433 | 2.876 | 3.395 | *5.366* | 2.530 | 3.768 | 2.634 | **1.879** |
| | -2 | Bias | **0.002** | -0.050 | -0.067 | *-0.374* | -0.056 | -0.048 | 0.019 | 0.186 |
| | | SE | 0.652 | 0.726 | 0.806 | *1.193* | 0.665 | 0.880 | 0.694 | **0.523** |
| | | MSE | 0.425 | 0.529 | 0.654 | *1.564* | 0.446 | 0.777 | 0.482 | **0.308** |
| | -1 | Bias | **0.001** | -0.023 | -0.030 | *-0.189* | -0.028 | -0.022 | 0.012 | 0.094 |
| | | SE | 0.329 | 0.368 | 0.412 | *0.629* | 0.336 | 0.447 | 0.351 | **0.263** |
| | | MSE | 0.108 | 0.136 | 0.171 | *0.431* | 0.114 | 0.200 | 0.123 | **0.078** |
| | 0 | Bias | **0.000** | **0.000** | **0.000** | *-0.007* | **0.000** | 0.001 | **0.000** | **0.000** |
| | | SE | **0.032** | 0.045 | 0.045 | *0.063* | **0.032** | 0.045 | **0.032** | **0.032** |
| | | MSE | **0.001** | 0.002 | 0.002 | *0.004* | **0.001** | 0.002 | **0.001** | **0.001** |
| | 2 | Bias | **-0.001** | 0.050 | 0.067 | -0.049 | 0.057 | 0.054 | -0.017 | *-0.188* |
| | | SE | 0.647 | 0.720 | 0.805 | *1.132* | 0.660 | 0.886 | 0.693 | **0.519** |
| | | MSE | 0.418 | 0.521 | 0.653 | *1.283* | 0.439 | 0.788 | 0.481 | **0.305** |
| | 5 | Bias | -0.045 | 0.067 | 0.088 | *-0.429* | 0.095 | **0.022** | -0.099 | *-0.492* |
| | | SE | 1.531 | 1.670 | 1.819 | *2.156* | 1.556 | 1.933 | 1.598 | **1.269** |
| | | MSE | 2.347 | 2.792 | 3.318 | *4.833* | 2.429 | 3.738 | 2.564 | **1.853** |

**Note:** SW: Shapiro-Wilk; AD: Anderson-Darling; CVM: Cramer-von Mises; PT: Pearson Chi-square; SF: Shapiro-Francia; LT: Lilliefors; JB: Jarque-Bera; AC: Artificial Covariate.

**Table 2. (Continuation)**

| N | true λ | | SW | AD | CVM | PT | SF | LT | JB | AC |
|---|---|---|---|---|---|---|---|---|---|---|
| | | Bias | **-0.009** | -0.097 | -0.109 | *-0.334* | -0.129 | -0.052 | 0.050 | 0.303 |
| | -5 | SE | 1.274 | 1.427 | 1.570 | *2.046* | 1.295 | 1.694 | 1.315 | **1.102** |
| | | MSE | 1.622 | 2.045 | 2.476 | *4.297* | 1.694 | 2.874 | 1.733 | **1.307** |
| | | Bias | **-0.001** | -0.041 | -0.049 | *-0.208* | -0.049 | -0.038 | 0.021 | 0.126 |
| | -2 | SE | 0.516 | 0.588 | 0.660 | *0.963* | 0.525 | 0.725 | 0.537 | **0.441** |
| | | MSE | 0.266 | 0.348 | 0.438 | *0.971* | 0.278 | 0.527 | 0.289 | **0.210** |
| | | Bias | **0.004** | -0.014 | -0.017 | *-0.099* | -0.020 | -0.016 | 0.016 | 0.067 |
| | -1 | SE | 0.253 | 0.289 | 0.325 | *0.485* | 0.258 | 0.360 | 0.264 | **0.218** |
| 50 | | MSE | 0.064 | 0.084 | 0.106 | *0.245* | 0.067 | 0.130 | 0.070 | **0.052** |
| | | Bias | **0.000** | 0.001 | **0.000** | *-0.002* | **0.000** | 0.001 | **0.000** | **0.000** |
| | 0 | SE | **0.032** | **0.032** | **0.032** | *0.045* | **0.032** | **0.032** | **0.032** | **0.032** |
| | | MSE | **0.001** | **0.001** | **0.001** | *0.002* | **0.001** | **0.001** | **0.001** | **0.001** |
| | | Bias | **-0.002** | 0.036 | 0.044 | -0.020 | 0.045 | 0.035 | -0.024 | *-0.129* |
| | 2 | SE | 0.513 | 0.586 | 0.659 | *0.929* | 0.522 | 0.720 | 0.533 | **0.437** |
| | | MSE | 0.263 | 0.345 | 0.436 | *0.864* | 0.275 | 0.520 | 0.285 | **0.208** |
| | | Bias | **0.010** | 0.107 | 0.124 | *-0.212* | 0.128 | 0.060 | -0.046 | *-0.311* |
| | 5 | SE | 1.286 | 1.441 | 1.587 | *1.991* | 1.308 | 1.699 | 1.324 | **1.107** |
| | | MSE | 1.655 | 2.089 | 2.534 | *4.008* | 1.726 | 2.889 | 1.755 | **1.322** |
| | | Bias | **-0.017** | -0.064 | -0.070 | *-0.179* | -0.108 | -0.057 | 0.030 | 0.171 |
| | -5 | SE | 0.964 | 1.133 | 1.264 | *1.751* | 0.978 | 1.367 | 0.986 | **0.875** |
| | | MSE | 0.929 | 1.288 | 1.603 | *3.098* | 0.968 | 1.871 | 0.973 | **0.795** |
| | | Bias | **0.000** | -0.018 | -0.019 | -0.072 | -0.036 | -0.014 | 0.019 | *0.076* |
| | -2 | SE | 0.387 | 0.452 | 0.505 | *0.741* | 0.393 | 0.552 | 0.395 | **0.351** |
| | | MSE | 0.150 | 0.205 | 0.255 | *0.555* | 0.156 | 0.305 | 0.156 | **0.129** |
| | | Bias | **0.001** | -0.009 | -0.010 | -0.035 | -0.017 | -0.007 | 0.011 | *0.039* |
| | -1 | SE | 0.192 | 0.223 | 0.251 | *0.368* | 0.194 | 0.274 | 0.195 | **0.172** |
| 100 | | MSE | 0.037 | 0.050 | 0.063 | *0.137* | 0.038 | 0.075 | 0.038 | **0.031** |
| | | Bias | **0.000** | **0.000** | **0.000** | *-0.001* | **0.000** | **0.000** | **0.000** | **0.000** |
| | 0 | SE | **0.000** | **0.000** | **0.000** | *0.032* | **0.000** | *0.032* | **0.000** | **0.000** |
| | | MSE | **0.000** | **0.000** | **0.000** | *0.001* | **0.000** | *0.001* | **0.000** | **0.000** |
| | | Bias | **0.008** | 0.029 | 0.033 | -0.011 | 0.044 | 0.030 | -0.012 | *-0.068* |
| | 2 | SE | 0.390 | 0.458 | 0.513 | *0.733* | 0.396 | 0.563 | 0.397 | **0.354** |
| | | MSE | 0.152 | 0.211 | 0.264 | *0.537* | 0.159 | 0.318 | 0.158 | **0.130** |
| | | Bias | **0.006** | 0.063 | 0.070 | -0.082 | 0.096 | 0.049 | -0.041 | *-0.185* |
| | 5 | SE | 0.972 | 1.142 | 1.273 | *1.722* | 0.987 | 1.382 | 0.993 | **0.882** |
| | | MSE | 0.945 | 1.308 | 1.626 | *2.973* | 0.984 | 1.912 | 0.988 | **0.812** |

**Note:** SW: Shapiro-Wilk; AD: Anderson-Darling; CVM: Cramer-von Mises; PT: Pearson Chi-square; SF: Shapiro-Francia; LT: Lilliefors; JB: Jarque-Bera; AC: Artificial Covariate.

**Table 2. (Continuation)**

| N | true λ | | SW | AD | CVM | PT | SF | LT | JB | AC |
|---|---|---|---|---|---|---|---|---|---|---|
| | | Bias | -0.009 | -0.015 | -0.019 | **0.006** | -0.048 | -0.012 | 0.010 | *0.049* |
| | -5 | SE | 0.518 | 0.628 | 0.693 | *1.002* | 0.521 | 0.757 | 0.520 | **0.500** |
| | | MSE | 0.268 | 0.394 | 0.480 | *1.004* | 0.274 | 0.573 | 0.271 | **0.252** |
| | | Bias | -0.004 | -0.005 | -0.006 | -0.004 | *-0.020* | -0.006 | **0.003** | 0.019 |
| | -2 | SE | 0.205 | 0.251 | 0.277 | *0.405* | 0.206 | 0.300 | 0.205 | **0.197** |
| | | MSE | 0.042 | 0.063 | 0.077 | *0.164* | 0.043 | 0.090 | 0.042 | **0.039** |
| | | Bias | **-0.001** | **-0.001** | **-0.001** | 0.002 | -0.009 | **0.001** | 0.003 | *0.011* |
| | -1 | SE | 0.105 | 0.126 | 0.138 | *0.202* | 0.104 | 0.148 | 0.105 | **0.099** |
| | | MSE | 0.011 | 0.016 | 0.019 | *0.041* | 0.011 | 0.022 | 0.011 | **0.010** |
| 500 | | Bias | **0.000** | 0.000 | 0.000 | 0.000 | 0.000 | 0.000 | 0.000 | 0.000 |
| | 0 | SE | **0.000** | 0.000 | 0.000 | 0.000 | 0.000 | 0.000 | 0.000 | 0.000 |
| | | MSE | **0.000** | 0.000 | 0.000 | 0.000 | 0.000 | 0.000 | 0.000 | 0.000 |
| | | Bias | 0.003 | 0.002 | 0.002 | -0.010 | 0.019 | **0.000** | -0.005 | *-0.020* |
| | 2 | SE | 0.205 | 0.251 | 0.277 | *0.406* | 0.206 | 0.302 | 0.207 | **0.199** |
| | | MSE | 0.042 | 0.063 | 0.077 | *0.165* | 0.043 | 0.091 | 0.043 | **0.040** |
| | | Bias | 0.005 | 0.006 | 0.008 | -0.031 | 0.045 | **0.003** | -0.015 | *-0.053* |
| | 5 | SE | 0.511 | 0.623 | 0.689 | *1.014* | 0.515 | 0.747 | 0.515 | **0.493** |
| | | MSE | 0.261 | 0.388 | 0.475 | *1.030* | 0.267 | 0.558 | 0.265 | **0.246** |

**Note:** SW: Shapiro-Wilk; AD: Anderson-Darling; CVM: Cramer-von Mises; PT: Pearson Chi-square; SF: Shapiro-Francia; LT: Lilliefors; JB: Jarque-Bera; AC: Artificial Covariate.

**Table 3. Results on textile data**

|         | SW     | AD     | CVM    | PT    | SF     | LT     | JB     | AC     |
|---------|--------|--------|--------|-------|--------|--------|--------|--------|
| $\hat{\lambda}$ | -0.060 | -0.080 | -0.100 | 0.020 | -0.060 | -0.060 | -0.060 | -0.044 |
| SW-pval | 1.000  | 1.000  | 1.000  | 1.000 | 1.000  | 1.000  | 1.000  | 1.000  |
| SF-pval | 1.000  | 1.000  | 1.000  | 1.000 | 1.000  | 1.000  | 1.000  | 1.000  |
| JB-pval | 1.000  | 1.000  | 1.000  | 1.000 | 1.000  | 1.000  | 1.000  | 1.000  |

Note: $\hat{\lambda}$ is the estimate of power transformation parameter, pval is the p-value of the corresponding test after transformation.

**Table 4. Results on students' grades data**

|         | SW    | AD    | CVM   | PT    | SF    | LT    | JB    | AC    |
|---------|-------|-------|-------|-------|-------|-------|-------|-------|
| $\hat{\lambda}$ | 1.910 | 1.760 | 1.580 | 1.270 | 1.970 | 1.540 | 1.780 | 1.393 |
| SW-pval | 1.000 | 1.000 | 0.804 | 0.152 | 1.000 | 0.700 | 1.000 | 0.345 |
| SF-pval | 1.000 | 1.000 | 0.804 | 0.152 | 1.000 | 0.700 | 1.000 | 0.345 |
| JB-pval | 1.000 | 1.000 | 1.000 | 0.269 | 1.000 | 1.000 | 1.000 | 0.616 |

Note: $\hat{\lambda}$ is the estimate of power transformation parameter, pval is the p-value of the corresponding test after transformation.

**Table 5. Computational Times (in seconds)**

|         | ALL  | SW   | AD   | CVM  | PT   | SF   | LT   | JB   | AC   |
|---------|------|------|------|------|------|------|------|------|------|
| textile | 2.55 | 0.28 | 0.33 | 0.32 | 0.25 | 0.31 | 0.34 | 0.20 | 1.68 |
| grades  | 2.65 | 0.30 | 0.36 | 0.39 | 0.25 | 0.31 | 0.33 | 0.22 | 1.67 |